\documentstyle[aps,prl,epsfig]{revtex}
\begin{document}
\title{Structure and Superconducting Properties of ``$\rm BeB_2$''}
\draft
\author{D.P. Young and P.W. Adams}
\address{Department of Physics and Astronomy\\Louisiana State University\\Baton Rouge, Louisiana,
70803}
\author{Julia Y. Chan and Frank R. Fronczek}
\address{Department of Chemistry\\Louisiana State University\\Baton Rouge, Louisiana,
70803}
\date{\today}
\maketitle

\begin{abstract}
We present the crystal structure and low temperature electronic transport properties of the
intermetallic commonly known as $\rm BeB_2$.  In contrast to the much simpler $AlB_2$-type
structure of the 39K superconductor $\rm MgB_2$, $\rm BeB_2$ forms a complex structure type that
is nearly unique in nature.  The structure has 110.5 atoms per unit cell and a stoichiometry
$\rm BeB_{2.75}$.  Polycrystalline $\rm Be(^{10.8}B)_{2.75}$ is superconducting below $T_c=0.72$K
with a critical magnetic field $H_{c2}=0.175$T.  Isotopically pure $\rm ^{10.0}B$ samples have
an enhanced $T_c=0.79$K.  Hall effect measurements suggest that the material is intrinsically
compensated. 
\\     
\end{abstract}
	Over the last few years there has been a renewed interest in intermetallic superconductors that
incorporate relatively low mass elements.  In the framework of BCS theory the lighter elements
result in higher frequency phonon modes and may thereby produce enhanced transition temperatures.
This idea is particularly well demonstrated in pure, disordered $\rm Be$ films, which have the
highest known elemental superconducting transition temperature
$T_c\approx10$K \cite{Be}.  Studies of intermetallics containing boron have been particularly
interesting and fruitful \cite{BC}, culminating in the recent discoveries of 39K and
9.5K superconductivity in the astonishingly simple binaries $\rm MgB_2$ \cite{MdB} and $\rm TaB_2$
\cite{TdB}, respectively.  Naively one would hope to obtain even higher transition temperatures by
substituting Be for the Mg in $\rm MgB_2$ given that the atomic mass ratio of Mg to Be is
$M_{Mg}/M_{Be}=2.25$.  This obvious substitution was almost certainly tried by many of the groups
studying $\rm MgB_2$ and unfortunately proved unsuccessful.  Indeed, Felner \cite{Felner} has
recently reported susceptibility measurements of $\rm BeB_2$ that show no sign of
superconductivity down to 5K.  Recent band structure calculations of $\rm BeB_2$, in which the
$AlB_2$-type structure of $\rm MgB_2$ was assumed, indicate that the Fermi surface topology of
$\rm BeB_2$ is sufficiently modified from that of $\rm MgB_2$ so as to preclude the possibility of
medium-$T_c$ superconductivity
\cite{Band}.  As a consequence of these findings the superconducting community has abandoned
its initial interest in $\rm BeB_2$.  In this Letter we demonstrate that the compound commonly
referred to as $\rm BeB_2$ is much more interesting than expected.  Our single crystal X-ray
refinement of the structure not only shows that the
$\rm BeB_2$ stoichiometry does not exist but that the correct stoichiometry, $\rm BeB_{2.75}$, is
{\it not} isostructural with $\rm MgB_2$.  Indeed, $\rm BeB_{2.75}$ forms a beautiful and
surprisingly complex crystal structure, see Fig.\ 1, that is almost unique in nature - only two
other compounds are known to have the same structure \cite{TwoComp}.  To our knowledge all
previous electronic band calculations \onlinecite{Band} of ``$\rm BeB_2$'' have assumed an
$AlB_2$-type structure and are therefore incorrect.  Interestingly, we have discovered that
$\rm BeB_{2.75}$ superconducts at a transition temperature $T_c=0.72$K.  Though this is somewhat
low by the standards of $\rm MgB_2$, it is still very promising in that some obvious schemes to
increase $T_c$, such as doping, may emerge once the electronic structure of the material is known.

 Though beryllium diboride has been known and studied for more than 40 years, there has
been no firm evidence that this phase exists with stoichiometry $\rm BeB_2$. The space group and
cell dimensions of a compound purported to be $\rm BeB_2$ were reported \cite{BeB2Struc} to be
hexagonal P6/$mmm$, a = 0.979(2) nm, c = 0.955(2) nm, but a full structure determination was not
carried out.  In the present study single crystals of "$\rm BeB_2$" were formed by slowly cooling
a $Be$-rich arcmelted boule of elemental Be and B.  The crystals formed as small thin plates
with a dull metallic luster and were mechanically extracted from the melt.  Single-crystal
intensity data were measured at 120 K on a Nonius KappaCCD diffractometer, and refinement with
606 reflections yielded a R=0.047.  Our refinement resulted in a stoichiometry of $\rm
Be_{1.23}B_{3.38}$ (i.e., $\rm BeB_{2.75}$) with a = 0.97738(7)nm and c = 0.95467(6)nm (P6/$mmm$,
Z = 24).

	The calculated structure is shown in Fig.\ 1a.  It is constructed from stacked hexagonal layers
of boron atoms, hexagons formed by B and Be, equilateral triangles of boron atoms, $\rm B_{12}$
icosahedra, and isolated Be atoms as shown in Fig.\ 1b.   In a manner similar to what is
observed in the newly-discovered superconductor $\rm MgB_2$ \onlinecite{MdB}, hexagonal layers of
B are stacked along the c axis with hexagonal layers of sites half-occupied by boron and
beryllium, as shown in Fig.\ 1c. The B-B bond distances are 0.18414(19)nm, and the B-Be
distances are 0.2016(2)nm. However, the layering is more complex than in $\rm MgB_2$, as
equilateral triangles formed by another layer of B, disordered around a sixfold axis, are also
present. The interleaved stacks, Fig.\ 1c, described above lie in channels formed by hexagonal
networks of $\rm B_{12}$ icosahedra, linked at the vertices by B-B bonds. These hexagonal networks
of $\rm B_{12}$ icosahedra are in turn linked along the c-axis direction by non-icosahedral $\rm
B_{12}$ cage polyhedra. The observed B-B distances of 0.17715(2)nm to 0.1820(2)nm in the
icosahedra are similar to those in the $\rm B_{12}$ icosahedra in
$\rm K_2B_{12}H_{12}$ (0.1775 nm) \cite{KBH} and in tetragonal boron (0.1806 nm) \cite{Icoso}. 
Be atoms lie in interstitial sites within the boron framework, including one Be site that was
treated as 1/2 occupied on a site of $mm$ symmetry, and another which was treated as 1/8
occupied on a site of $3m$ symmetry.  More details of the crystal structure will be reported
elsewhere \cite{BeB275}. 

	Low temperature magnetotransport studies were made on samples cut from polycrystalline
boules of $\rm BeB_{2.75}$.  The resistance of rectangular samples of typical dimensions 
2mm x 0.5mm x 0.1mm was measured using a standard 4-probe ac technique.  The samples were cooled
in a dilution refrigerator down to 50mK in magnetic fields up to 6T.  The polycrystalline samples
had room temperature resistivities of $\sim3000\mu\Omega$-cm that changed very little upon
cooling, falling only $\sim10\%$ from room temperature to 4K.  The
resistivity remained essentially unchanged at lower temperatures.  We made Hall effect
measurements up to 6T at 100mK in order to determine the carrier density $N$ but could not extract
a measurable Hall voltage.  Assuming single carrier conduction, our Hall measurements
indicate a lower bound on the carrier density of $N>10^{22}cm^{-3}$.  This limit is inconsistent
with the high resistivity of our samples, suggesting that $\rm BeB_{2.75}$ may be intrinsically
compensated. 

 Shown in Fig.\ 2 are the resistivities as a function of temperature of two samples, one made from
natural abundance $\rm ^{10.8}B$ and the other from isotopically pure $\rm ^{10}B$.  It is evident
from Fig.\ 2 that $\rm BeB_{2.75}$ superconducts with $T_c(^{10.8}\rm B)=0.72$K and that the $\rm
^{10}B$ sample has a slightly enhanced $T_c(^{10.0}\rm B)=0.79$K.  If the boron phonon
modes are soley responsible for mediating the superconductivity, then BCS theory predicts that
the $^{10}\rm B$ transition temperature should be enhanced by the factor ${(10.8/10.0)}^\alpha$
where $\alpha=1/2$ \cite{Tinkham,Canfield}.  $T_c(^{10.0}\rm B)$ in Fig.\ 2 is about twice as
large as the expected value $T_c(^{10.0}\rm B)\sim(1.04) T_c(^{10.8}\rm B)$, suggesting that
$\alpha\sim1$. The deviation from $\alpha=1/2$ may be due to the complexity of the BCS state in
this material.  It is also possible that it is representative of non-phonon mediated
superconducitivity \cite{Hirsch}.

 The open symbols in Fig.\ 2 were data taken in a field of 1T which is
above the upper critical field $H_{c2}\approx0.18$T.  Note that the normal state resistivity in
Fig.\ 2 is temperature independent.  We also found the normal state to be magnetic field
independent below 1K.  This leads us to believe that the relatively high resistivity of our
samples is a consequence of a low carrier concentration and not disorder. In Fig.\ 3 we show the
resistive upper critical field transitions at several temperatures of the $\rm ^{10}B$ sample. 
The increase of the transition width with decreasing temperature (see inset of Fig.\ 4) is
indicative of flux flow broadening.  This suggests that the $\rm BeB_{2.75}$ is a type II
superconductor, as is $\rm MgB_2$ \cite{Canfield2}.  At $T=100$mK the upper critical
field is $H_{c2}=0.19$T, which gives a coherence length $\xi_o\sim30$nm. In Fig.\ 4 we plot the
upper critical field, as measured from the mid-points of the transitions in Fig.\ 3, as a
function of the reduced temperature $1-T/T_c$. Note the extreme linearity of the data, in
contrast to the more typical quadratic dependence seen in many BCS superconductors
\onlinecite{Tinkham}.  The solid line in Fig.\ 4 is a linear least-squares fit to the data points
near the origin, and has a slope $dH_{c2}/dT\approx0.2$T/K.  Similar linearity was also obtained
from critical fields determined by an onset criterion.

	In conclusion, we have refined the crystal structure of ``$\rm BeB_2$'' from single-crystal
X-ray diffraction data and find that not only is $\rm BeB_2$ an unstable stoichiometry but
that the correct stoichiometry $\rm BeB_{2.75}$ has an extremely complex and rare crystalline
structure type.  Though the superconducting transition temperature of $\rm BeB_{2.75}$ is
relatively low, one must keep in mind that virtually nothing is known about its electronic band
structure.  As has been the case with perovskites \cite{HTS,Cava}
and the $\rm C_{60}$ fullerenes \cite{C60}, the rich structural framework of $\rm BeB_{2.75}$ may
afford opportunities to increase $T_c$ once we have better understanding of its electronic
properties.  

	We gratefully acknowledge useful discussions with John DiTusa, Richard Kurtz, Art Ramirez,
Vladimir Butko, Roy Goodrich and Dana Browne.  This work was supported by the NSF under Grant No.
DMR 99-72151.


\newpage

%

\begin{figure}
\centerline {FIGURES}
\vspace{0.5in}
\centerline{\epsfig{file=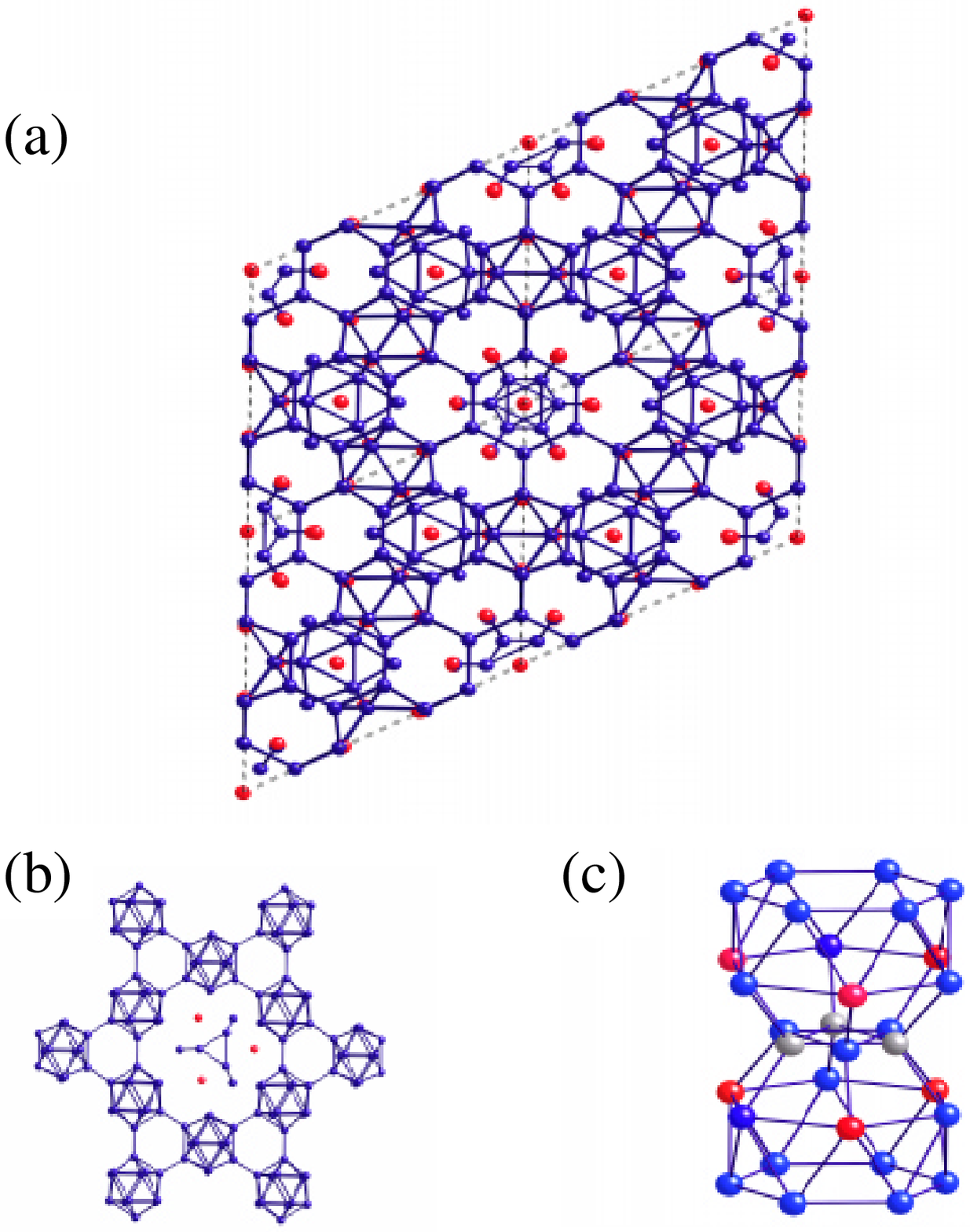,width=6.0in}}
\vspace{0.6in}
\caption{a) Four unit cells of the $\rm BeB_{2.75}$ structure are shown, where Be and B atoms
are represented in red and blue, respectively.  Note: The hexagonal cluster at the center of the
figure is shown in greater detail in panel c). b) The hexagonal network of vertex-linked
icosahedra is shown along the c direction.  c) Layers of B, B/Be hexagons, and the disordered
equilateral B triangles form a building block of the structure. Only one of the two possible
orientations of the channel contents is shown.}
 \label{Figure 1}
 \end{figure}
\newpage

\begin{figure}
\vspace{0.5in}
\centerline{\epsfig{file=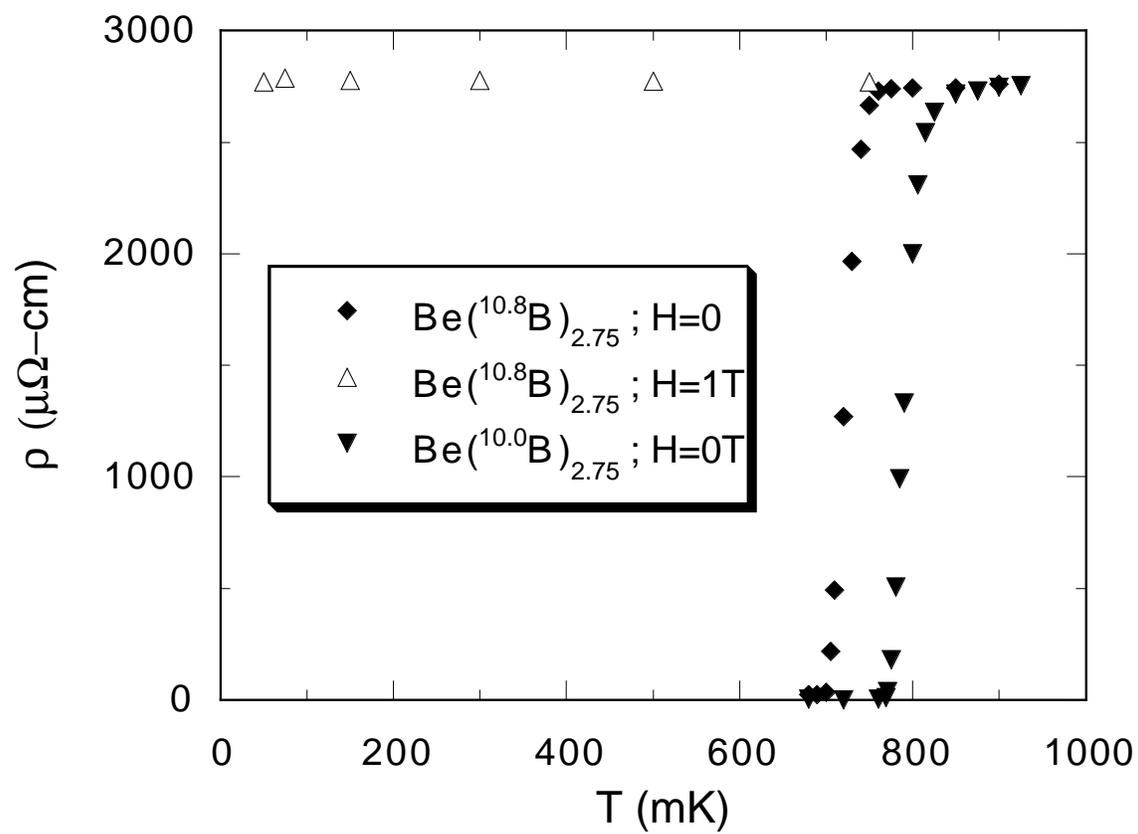,width=6.0in}}
\vspace{0.6in}
\caption{Sample resistivity as a function of temperature. 
Solid symbols: zero field.  Open symbols: H=1T. The open symbols represent the normal state.}
 \label{Figure 2}
 \end{figure}
\newpage

\begin{figure}
 \vspace{02.5in}
\centerline{\epsfig{file=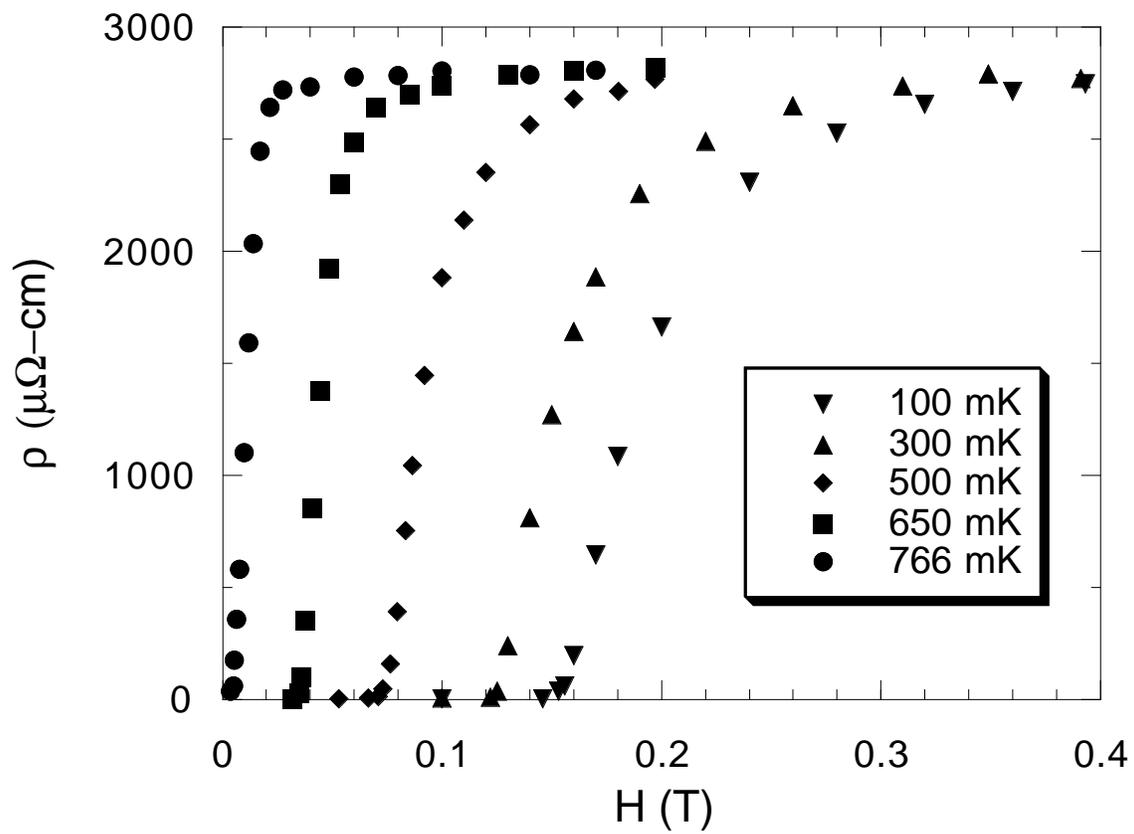,width=6.0in}}
\vspace{0.3in}
\caption{Resistive upper critical field transitions for the $\rm ^{10}B$ sample.}
 \label{Figure 3}
 \end{figure}
\newpage

\begin{figure}
\vspace{2.5in}
\centerline{\epsfig{file=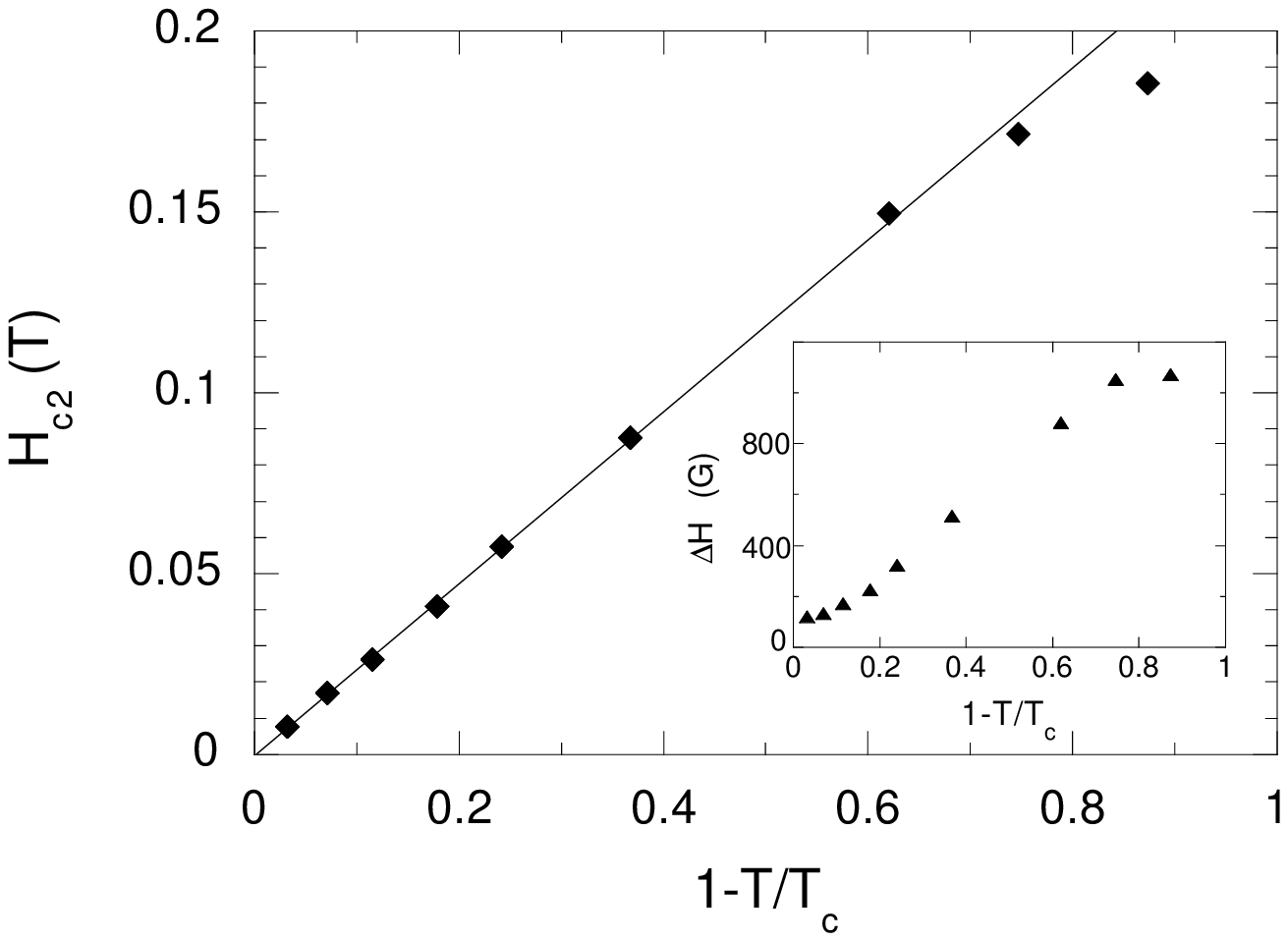,width=6.0in}}
\vspace{0.3in}
\caption{Upper critical field as a function of reduced temperature.  $H_{c2}$ was taken as the
midpoints of the transitions in Fig.\ 3. The solid line is least-squares fit to the data points
near the origin. Inset: Critical field ($10\% - 90\%$) width as function of reduced temperature.}
 \label{Figure 4}
 \end{figure}
\newpage

%

%


\begin{references}

\bibitem{Be} B.G. Lazarev, A.I. Sudovtsov, and A.P. Smirnov, Zh. Eksp. Teor. Fiz. 33, 1059
(1957) [JETP {\bf6}, 816-817 (1958)]; K. Takei, K. Nakamura, and Y. Maeda, J. Appl. Phys. 57 (11),
5093-5094 (1985).

\bibitem{BC} R. Nagarajan, {\it et al.}, Phys. Rev. Lett. {\bf72}, 274 (1994); R.J.
Cava, {\it et al.}, Nature (London) {\bf376}, 146 (1994).

\bibitem{MdB} J. Nagamatsu, N. Nakagawa, T. Muranaka, Y. Zenitani, and J. Adimitsu, Nature
{\bf410}, 63 (2001).

\bibitem{TdB} D. Kaczorowski, A.J. Zaleski, O.J. Zogal, and J. Klamut, cond-mat/0103571 v2.

\bibitem{Felner} I. Felner, cond-mat/0102508.


\bibitem{Band} N.I. Medvedeva, A.L. Ivanovskii, J.E. Medvedeva, and A.J. Freeman,
cond-mat/0103157.

\bibitem{TwoComp}  The other two compounds believed to have the structure of Fig.\ 1 are
ternaries, $\rm AlB_{51}Be_{17}$ and $\rm AlB_{57}Be_{17}$.  The structure type is named after the
latter compound. V.R. Mattes, {\it et al.}, Z Anorg. Allg. Chem. {\bf413}, 1 (1975).

\bibitem{BeB2Struc} D.E. Sands, {\it et al.}, Acta. Crystallogr. {\bf14}, 309 (1961); C.L.
Hoenig, C.F. Cline, and D.E. Sands, J. Amer. Ceram. Soc. {\bf44}, 385 (1961).

\bibitem{KBH} J.A. Wunderlich and W.N. Lipscomb, J. Am. Chem. Soc. {\bf}, 4427 (1960).

\bibitem{Icoso} J.L. Hoard, R.E. Hughes, and D.E. Sands, J. Am. Chem. Soc. {\bf80}, 4507 (1958).

\bibitem{BeB275} J.Y. Chan, F.R. Fronczek, D.P. Young, and P.W. Adams, in preparation.

\bibitem{Tinkham} M. Tinkam, {\it Introduction to Superconductivity}, 2nd ed., (McGraw-Hill, New
York,1996).

\bibitem{Canfield} A boron isotope effect has also been reported in $MgB_2$.  S.L. Bud'ko, {\it et
al.}, Phys. Rev. Lett. {\bf86}, 2423 (2001).


\bibitem{Hirsch} J.E. Hirsch, cond-mat/0102115.

\bibitem{Canfield2} D.K. Finnemore, {\it et al.}, Phys. Rev. Lett. {\bf86}, 2420 (2001).


\bibitem{HTS} J.G. Bednorz and K.A. Muller, Rev. Mod. Phys. {\bf60} (3), 585 (1988).

\bibitem{Cava} T.He. Huang, {\it et al.}, cond-mat/0103296.

\bibitem{C60} A.F. Hebard, {\it et al.}, Nature {\bf350}, 600 (1991); O. Gunnarsson, Rev. Mod.
Phys. {\bf69} (2), 575 (1997).


\end{references}
\end{document}